# Effect of social isolation in dengue cases in the state of Sao Paulo, Brazil: an analysis during the COVID-19 pandemic

Gleice Margarete de Souza Conceição[1], Gerson Laurindo Barbosa[2], Camila Lorenz*[1], Ana Carolina Dias Bocewicz[3], Lidia Maria Reis Santana[4,5], Cristiano Corrêa de Azevedo Marques[2] & Francisco Chiaravalloti-Neto[1]

1 Department of Epidemiology, School of Public Health, University of Sao Paulo, Sao Paulo, Brazil.

2 Endemics Control Superintendence (SUCEN), Sao Paulo State Department of Health, Sao Paulo, Brazil.

3 Disease Control Coordination (CCD SES/SP), Sao Paulo, Brazil.

4 Epidemiological Surveillance Center "Professor Alexandre Vranjac" - Sao Paulo State Department of Health (CVE/SES-SP), Sao Paulo, Brazil.

5 Federal University of São Paulo (UNIFESP), Sao Paulo, Brazil.

* Corresponding Author: C. Lorenz, Department of Epidemiology, School of Public Health, University of Sao Paulo, Brazil. E-mail: camilalorenz@usp.br

## ABSTRACT

**Background:** Studies have shown that human mobility is an important factor in dengue epidemiology. Changes in mobility resulting from COVID-19 pandemic set up a real-life situation to test this hypothesis. Our objective was to evaluate the effect of reduced mobility due to this pandemic in the occurrence of dengue in the state of São Paulo, Brazil. **Method:** It is an ecological study of time series, developed between January and August 2020. We use the number of confirmed dengue cases and residential mobility, on a daily basis, from secondary information sources. Mobility was represented by the daily percentage variation of residential population isolation, obtained from the Google database. We modeled the relationship between dengue occurrence and social distancing by negative binomial regression, adjusted for seasonality. We represent the social distancing dichotomously (isolation versus no isolation) and consider lag for isolation from the dates of occurrence of dengue. **Results:** The risk of dengue decreased around 9.1% (95% CI: 14.2 to 3.7) in the presence of isolation, considering a delay of 20 days between the degree of isolation and the dengue first symptoms. **Conclusions:** We have shown that mobility can play an important role in the epidemiology of dengue and should be considered in surveillance and control activities.

## 1. Introduction

Dengue, an acute disease caused by four virus serotypes belonging to the Flaviviridae family and transmitted mainly by *Aedes aegypti* mosquito, is the arbovirus that most affects the human population worldwide [1,2]. Approximately four billion people in 128 countries are at risk of infection [3], with the number of reported cases growing every year. For example, in 2019 alone, the World Health Organization recorded more than four million cases of dengue worldwide, 75% of which occurred in the Americas [4]. In Brazil, the first epidemic occurred in the 1980s and since then, cases have been increasing dramatically. In 2015 it reached the mark of 830 cases per 100 thousand inhabitants [5], being considered the largest epidemic in the country. In 2019 the dengue incidence rate reached 723 per 100 thousand inhabitants [6]. The most effective way to prevent outbreaks is to plan vector mosquito control strategies and minimize vector-to-human transmission, since vaccines against dengue (DENV) already developed or under development have not proven to be safe enough to be used routinely to control the disease [7].

Since *Ae. aegypti* rarely disperses over a distance of more than 100 meters from breeding grounds, these mosquitoes are prone to rest and carry out the blood meal inside houses, residential buildings, work and study places, among others [8,9,10]. Thus, human movements can be important to explain the dynamics of urban transmission of DENV [11,12,13]. Although human mobility can lead to transmission at multiple spatial and temporal scales [14, 15, 16], it is at the finer scales (daily intra-urban human movements) that most important epidemic processes and emergency public health interventions are generally implemented [2]. Empirical studies indicates that individual and spatial variability in frequency and number of contacts can lead to heterogeneous transmission, where some individuals or areas contribute disproportionately to the transmission of pathogens and, consequently, epidemic spread [17,18].

According to Barbosa et al. [19], most dengue cases occur in larger municipalities, considered to be the region's pole municipalities, that is, the occurrence of dengue cases is related to the proximity of these municipalities. This phenomenon also leads us to consider the influence of human mobility on dengue transmission. According to official records, more than 60% of dengue cases registered in the period from 2010 to 2020 occurred in municipalities with more than 100 thousand inhabitants, which represent approximately 70% of the state's population. Also according to Barbosa et al [19], the municipalities with the highest values of larval infestation presented more cases of dengue. However, these infestation indicators have not been able to indicate the chance of dengue occurrence, that is, it is not possible to measure the risk of cases from the mosquito's entomological indicators, since there are multiple factors involved in the occurrence of dengue.

The most populous State in the country, with approximately 45 million inhabitants, Sao Paulo is also highly interconnected to Brazil and the world. Its main airport, São Paulo-Guarulhos International Airport, is the largest in Brazil, with direct passenger flights to 103 destinations in 30 countries and 52 domestic flight routes [20]. São Paulo's status as an air traffic center can facilitate the rapid spread of viruses like DENV in the presence of the vector throughout our territory.

The current COVID-19 pandemic and social isolation measures to try to stem the spread of the virus have drastically reduced human mobility in some of the most critical periods of the disease. Thus, it is possible that the transmission of DENV has also been altered due to this reduction in human movements [21]. Therefore, the objective of this study was to assess the influence of reduced mobility due to the COVID-19 pandemic on the occurrence of dengue cases in 2020 in the State of São Paulo.

## 2. Materials and Methods

### 2.1 Study Area and Data Acquisition

The study had an ecological design of time series, covering the period from January 1 to August 11, 2020 (symptoms onset between 01-Jan-2020 and 11-Aug-2020, updated on 9-Sep-2020). The study area included the state of São Paulo, which registered a total of 183,792 dengue cases [22] and 749,059 COVID-19 cases during the study period [23]. It is located in the Southeast region of Brazil (Figure 1) and has a population of 44.6 million inhabitants, estimated for 2020, in a territorial area of 248,219 km², which corresponds to a demographic density of 180 inhabitants per km². The state capital is the municipality of São Paulo, with a population of 12.3 million inhabitants, the most populous municipality in the country. The State is home to 21% of the inhabitants of Brazil, its Gross Domestic Product in 2019 was R $ 2.1 trillion, corresponding to 29% of the Brazilian GDP, and has a Human Development Index (HDI) of 0.783, the best among all states in the country. Its climate can be characterized as tropical.

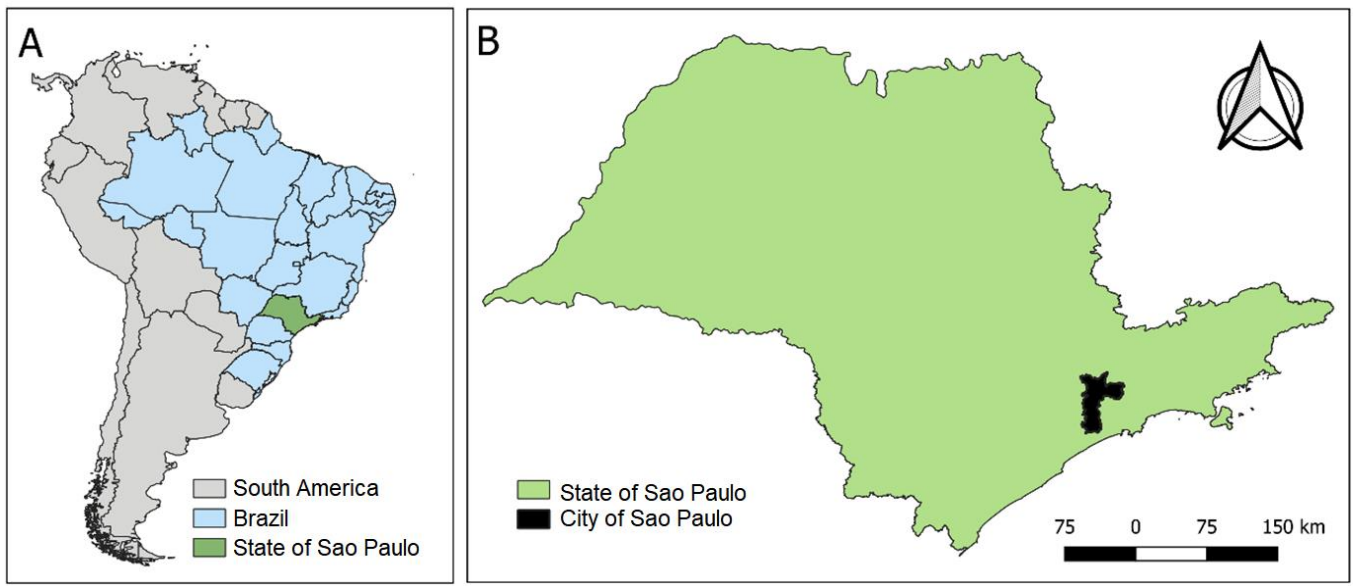

**Figure 1**. Study Area. **A.** South America, Brazil and the state of São Paulo. **B.** State of São Paulo and its Capital (municipality of São Paulo).

We use the number of confirmed dengue cases and residential mobility in the State of São Paulo in 2020, on a daily basis, from secondary sources of information. The number of confirmed dengue cases was obtained from the databases of the Epidemiological Surveillance Center of the São Paulo State Secretariat [22]. The daily percentage variation of residential population isolation was obtained from the Google mobility database - residential category, which collects mobility information, anonymously, from the cell phone of the population of users who activated the "Google Location History" setting. This percentage corresponds to the percentage variation of the population average of the time that users spent inside the home, from February 7, in relation to a baseline value, represented by the median observed on the same day of the week in the pre-epidemic period, of 3 from January to February 6, 2020 [24]. Negative values for this variable indicate that population mobility was greater than in the baseline period, positive values indicate lower mobility than baseline, that is, increased isolation.

Regarding the ethical aspects of the information used, it should be noted that the aggregated dengue data are in the public domain and are freely accessible on the websites of the São Paulo State Department of Health and the same applies to the Google platform mobility data. In this way, since it does not harm citizens' secrecy and confidentiality, it is exempted from analysis by an ethics committee in research on human beings.

## 2.2 Data Analysis

To evaluate the association between isolation and dengue occurrence, regression models with negative binomial response were used. The response variable was the daily number of dengue cases and the explanatory variables were a smoother for the number of cases, to control long-term seasonality (the natural spline function, with 8 nodes), indicators for the days of the week and the daily percentage of isolation. To control for the presence of serial autocorrelation, commonly found in time series, a first order autoregressive component was added to the model. Once the non-linear character of the relationship between the percentage of isolation and the number of cases was found, it was decided to represent isolation by means of a dichotomous categorical variable, with "no isolation" levels (for values between -4 and 0 %) and "with isolation" (for values between 1 and 30%).

It is reasonable to assume that there is a gap between the variation in isolation and the consequent impact on the number of dengue cases, that is, if an increase in isolation results in a decrease in the number of cases, this decrease does not occur on the same day, but some days after that increase. Thus, models were adjusted considering different lags for isolation: 5, 10, 15 and 20 days. The mean isolation was also considered in different periods prior to the onset of symptoms: in the last 7 days, between the 7th and the 14th day and between the 14th and the 21st day. All analysis and models were performed in R for Windows software.

## 3. Results

We found 185,806 confirmed dengue cases in the State of São Paulo between January 1st and August 11th of 2020 (829 cases per day, on average), the vast majority (72.9%) between January and March. From January 1st, the number of cases increased abruptly, reaching its peak on February 11th, when 2,620 cases were recorded. Thereafter, there was a decline until the end of the period (Figure 2). In the 11 days observed in the month of August, the daily average was 37 cases. The number of cases was slightly higher on Mondays than on other days of the week (Figure 3).

The daily percentage of isolation ranged from -4 to 30%. In the baseline period (January 3 to February 6), this variable assumed zero value. In most (57%) of the days between February 15 and March 13, isolation was negative, that is, the population spent more time away from home than in the baseline period. Thereafter, isolation was positive, ranging from 2 to 30%, with a peak on April 10 (15th epidemiological week) (Figure 2).

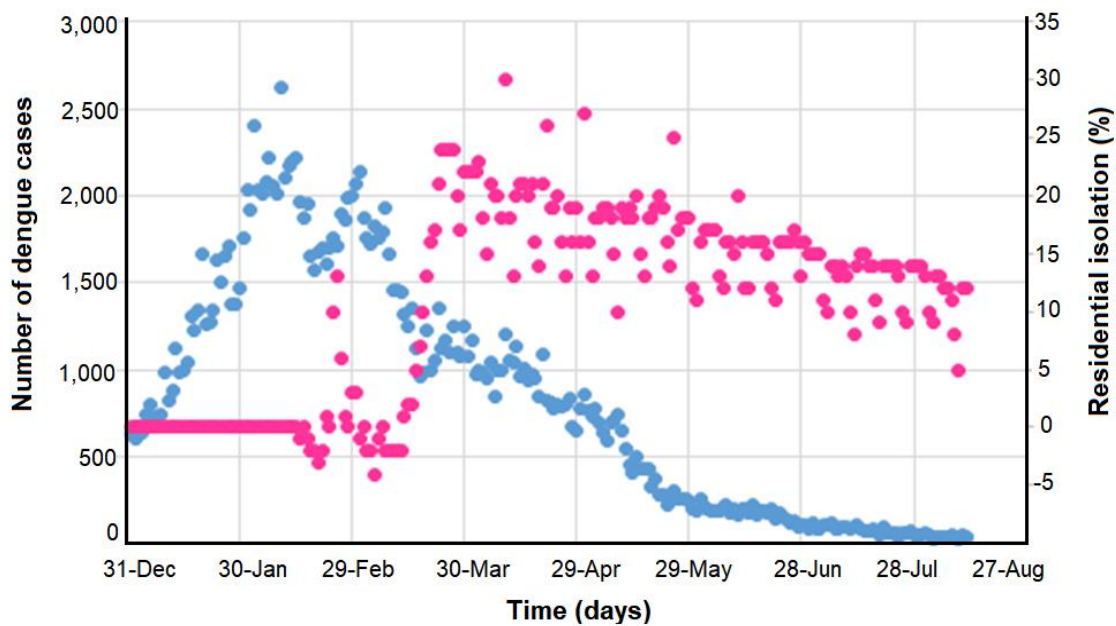

**Figure 2.** Number of dengue cases (blue, left) and percentage of daily isolation (pink, right) over time. State of São Paulo, January to August 2020.

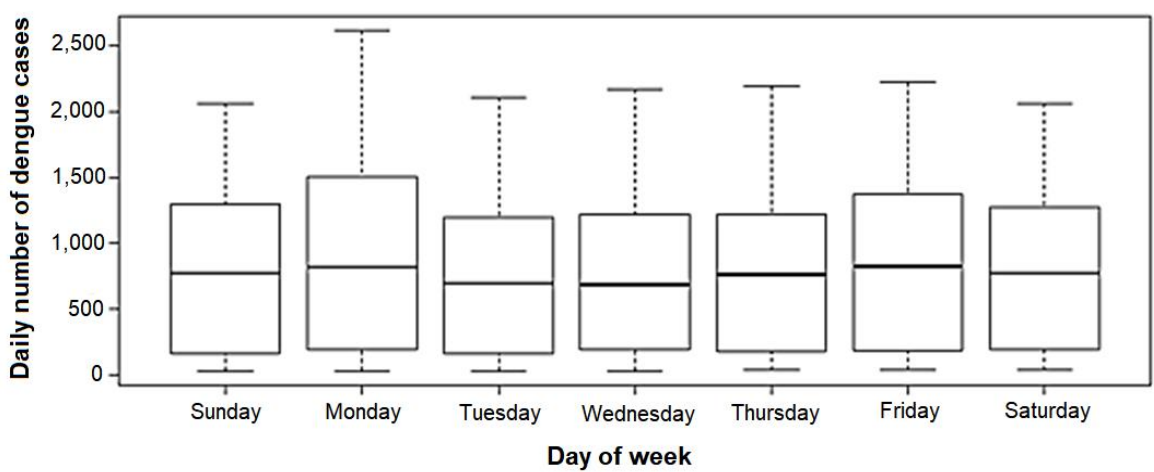

**Figure 3.** Distribution of the daily number of dengue cases according to the day of week. State of São Paulo, January to August 2020.

The number of cases showed a significant correlation with all isolation measures considered, with an inverse relationship, that is, the greater the isolation, the smaller the number of cases. The greatest correlation was for isolation 20 days before the onset of symptoms (Table 1). In general, the number of cases was higher on days when isolation was negative or zero than on days when it was positive. On these two occasions, the median number of cases was 1712 and 205, respectively (Figure 4).

**Table 1.** Spearman's correlation coefficients between daily number of dengue cases and daily measures of residential isolation.

| Isolation | Spearman |
|---|---|
| In the same day | -0,33* |
| In 5th day before | -0,39* |
| In 10th day before | -0,45* |
| In 15th day before | -0,51* |
| In 20th day before | -0,55* |
| 7-day moving average | -0,31* |
| Average between the previous 7th and 14th day | -0,40* |
| Average between the previous 14th and 21th day | -0,48* |

* *p-value* < 0,001

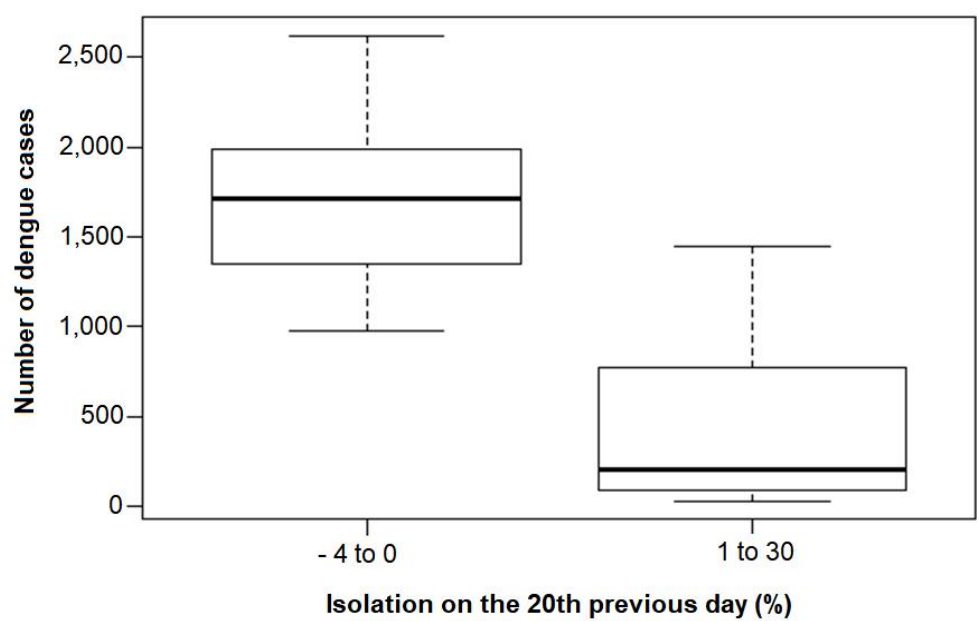

**Figure 4.** Distribution of the daily number of dengue cases according to residential isolation. State of São Paulo, January to August 2020.

Table 2 contains the estimates for the relative risk, as well as for the percentage variation of the risk of contracting dengue on days with isolation when compared to days without isolation in regression models with successive inclusion of control variables. In a model that contains only isolation as an explanatory variable, the RR was 0.24, which corresponds to a 75.6% decrease in risk on isolation days. The isolation effect decreased as control variables were included, but it remained significant, even after adjusting for the serial autocorrelation, using an order 1

8autoregressive term. The estimates for the complete model (the last model in the Table 2) are shown in table 3.

The RR was lower on all days of the week, when compared to Monday (p < 0.01 in all cases). According to the confidence intervals presented, there does not seem to be a difference between the RR of the other days. The risk of contracting dengue on days with isolation was 0.91 (CI = 0.86: 0.96) times the risk on days without isolation, that is, the risk decreases around 9.1% (CI = -14.2: -3.7) in the presence of isolation (Table 3).

**Table 2.** Relative risk and percentage variation of risk in the presence of isolation (delay of 20 days from the onset of symptoms of dengue cases) in regression models with successive inclusion of control variables. State of São Paulo, January to August 2020.

| Explanatory variables | RR[a] | CI 95% (RR) | PV(%)[b] | CI 95% (PV) | *p-value* |
|---|---|---|---|---|---|
| Isolation (lag 20 days) | 0.24 | (0.19 : 0.31) | - 75.6 | (- 80.9 : - 75.6) | < 0.001 |
| Isolation (lag 20 days) + seasonality | 0.90 | (0.82 : 0.98) | - 9.9 | (- 17.5 : - 9.9) | 0.02 |
| Isolation (lag 20 days) + seasonality + day of week | 0.88 | (0.82 : 0.95) | - 11.8 | (- 18.4 : - 11.8) | < 0.001 |
| Isolation (lag 20 days) + seasonality + day of week + first order autoregressive component | 0.91 | (0.86 : 0.96) | - 9.1 | (- 14.2 : - 3.7) | < 0.001 |

[a] Risk of contracting dengue in the isolation period (1 to 30%) when compared to the non-isolation period (-4 to 0%).
[b] (RR-1)*100: Percentage variation in the risk of contracting dengue due to isolation.

**Table 3.** Estimates of the final regression model for the occurrence of dengue cases, considering isolation with a 20-days delay from the onset of symptoms. State of São Paulo, January to August 2020.

| Explanatory variables | RR[a] | CI 95% (RR) | PV(%)[b] | CI 95% (PV) | *p-value* |
|---|---|---|---|---|---|
| **Day of week** | | | | | |
| Monday | 1.00 | | 0 | | |
| Tuesday | 0.84 | (0.81 : 0.88) | - 15.8 | (- 19.4 : - 12.0) | < 0.001 |
| Wednesday | 0.85 | (0.81 : 0.90) | - 14.8 | (- 19.5 : - 9.9) | < 0.001 |
| Thursday | 0.84 | (0.80 : 0.89) | - 15.5 | (- 20.2 : - 10.6) | < 0.001 |
| Friday | 0.93 | (0.88 : 0.98) | - 7.4 | (- 12.3 : - 2.2) | < 0.001 |
| Saturday | 0.88 | (0.83 : 0.92) | - 12.3 | (- 16.7 : - 7.7) | 0.005 |
| Sunday | 0.87 | (0.84 : 0.91) | - 12.6 | (- 16.1 : - 8.9) | < 0.001 |
| **Isolation** | | | | | |
| No (- 4% to 0) | 1.00 | | 0 | | |
| Yes (1 a 30%) | 0.91 | (0.86 : 0.96) | - 9.1 | (- 14.2 : - 3.7) | < 0.001 |

[+] Controlled for seasonality and autocorrelation.

## 4. Discussion

Our results showed an association between the presence of isolation and the occurrence of dengue in the state of São Paulo. The risk of occurrence of the disease was about 9% lower in the period in which there was isolation and remained significant even after the inclusion of successive control variables. Such variables allowed to take into account seasonality and the presence of autocorrelation frequently observed in daily data, providing more robust estimates. Several recent studies have shown the contribution of human mobility in the location and spatio-temporal evolution of the occurrence of dengue cases [2,13,25,26,27], either locally through long distance travel or migration. However, in general, the implications of spreading diseases are complex because contact between individuals is structured according to human activities, rather than assuming a random character. If hot spots of transmission and vector populations occur following some pattern, this can potentially provide targets for control interventions [26].

The characteristics of mosquito *Ae. aegypti* of biting during the day, usually inside the buildings and with reduced flight radius, make the transmission have a localized character. Considering that human mobility connects different areas, possibly with different population densities of mosquitoes, it can be deduced that the population's movement pattern is an important factor to define the dynamics of transmission and spread of dengue. According to Soriano-Panos et al [28], understanding the interdependence between human mobility and contagion processes is one of the keys to explain the onset and development of large-scale epidemics. Such knowledge can expand the ability to target areas of origin and areas where transmission can spread, or at least help to reduce costs associated with these efforts. For Schaber [2], there are two important conditions that must be considered in an epidemic: illness, which makes the person stay at home longer, increasing the chance of being bitten by a mosquito and maintaining local transmission around the environment; and cases of asymptomatic or oligosymptomatic individual, who continues to move and takes the disease to other areas, causing the cycle to repeat itself. Such conditions and their influence on both spread and disease control may vary based on population, cultural and socioeconomic distribution and density, that is, they may vary from city to city according to their structure [26] . It is important to consider that human mobility, especially of infected individuals, can create multiple waves of dengue, making it essential to consider different mobility models, since they can present different results [29].

The results found in our study showed a greater correlation for isolation 20 days before the onset of dengue symptoms, which is expected considering the disease cycle. Mosquitoes are infected by biting an infected human (viremic stage), and can then transmit the infection to others after an extrinsic incubation period (IPE) of 8 to 12 days [30]. After the intrinsic incubation period (IIP), which already occurs in the host and lasts approximately 4 to 15 days, dengue symptoms begin to manifest [31]. The adult vector *Ae. aegypti* survives on average for 30 days [32]. Consequently, the estimated lag time can be as short as 20 days if the dengue vectors bite a susceptible host in the first days after IPE. But it is important to remember that there are other variables that influence EIP and IIP, such as environmental conditions [33,34].

Unlike what we observed in the State of São Paulo, Nacher et al [35] showed that the number of dengue cases increased in French Guiana during social distance due to the COVID-19 pandemic, which occurred simultaneously with the rainy season. Therefore, it is difficult to separate the respective individual contributions from isolation and environmental conditions (temperature and rain) in the control of mosquitoes and dengue, since these factors interact with each other [35].

In this work we made several assumptions to simplify the analysis and the model built. We restricted the study to a Brazilian state and neglected spatial heterogeneities due to the lack of spatial data. In the future, more realistic simulations of dengue should include data on the distribution of the population within a city, the presence of potential mosquito breeding sites and other land cover characteristics that help to explain spatial heterogeneities. The inclusion of environmental variables, such as temperature and precipitation, can also serve as a proxy to explain the vector's reproduction time and the frequency of dengue cases. However, our study is one of the first to use human mobility data on a daily scale and associate them with the number of dengue cases, but it is certainly not an exhaustive exploration of the implications of human mobility patterns for the control of this disease.

Another point that deserves to be highlighted is that the reduction in the registration of dengue cases during the quarantine period may have been influenced by underreporting, as suggested by others [36]. The mobilization of epidemiological surveillance teams to respond to the emergence of the COVID-19 pandemic after the confirmation of the first cases in Brazil may have caused delay or underreporting of dengue cases. In addition, most dengue cases have symptoms that overlap with COVID-19 cases, sharing clinical characteristics [20].

Despite the various limitations to which the results and conclusions of this study are subject, its strongest point lies in the opportunity to test the hypothesis of the influence of mobility on the occurrence of dengue in a real-life situation. This was made possible by the COVID-19 pandemic experienced in the study area and across the planet, which has social isolation as one of the main control measures. The restrictions on people's displacement and mobility created a favorable situation to test the study's hypothesis. Even if simple modeling only partially covers complex problems such as the occurrence of dengue, advancing the understanding of the role of mobility in the epidemiology of this condition, made possible by the present study, should be a goal to be pursued [37]. Another strength of the study refers to the use of adequate statistical modeling to assess the relationship between dengue and mobility restriction, which takes into account the influence of seasonal factors inherent to the studied phenomenon, as well as lags between isolation and the occurrence of cases.

The current situation of the disease, with repeated seasonal epidemics, suggests the need for changes in the strategies currently used and the improvement of surveillance actions [38,39]. Some studies point to integrated control as the best alternative, but reveal major deficiencies in the areas of management, entomology, communication and financing [40]. In addition, particularly for the dengue control program, there is evidence that indicates low coherence and small scope achieved in relation to general and specific objectives, indicating the need to update the theoretical-logical model at different levels of management [41]. Despite the continuous activities implemented by the State of São Paulo and its municipalities, the historical series shows a cyclical increase in incidence, associated with a constant rate of infestation over the last 30 years [42]. In this context, actions directed exclusively to the vector are verified, and human mobility has been considered little or nothing as a relevant element in control strategies until now.

Based on our results, new studies on the influence of mobility on the occurrence of dengue cases may be developed, on a regional basis, from different units of analysis such as municipalities or other types of territorial organization (intra or interregional), both in the State of São Paulo and in other Brazilian territories. Such analyzes may provide particular information to certain local characteristics which, if taken into account, may assist managers in making decisions.

## 5. Conclusions

By establishing a time lag between the occurrence of dengue and social isolation, adjusting for seasonality, it was possible to show the inverse relationship between the two variables. Considering a 20-day lag between the degree of isolation and the occurrence of the disease, it was found that the risk of contracting dengue decreased around 9.1% in the presence of isolation compared to the days without it. This result showed that mobility can play an important role in the epidemiology of dengue and should be a factor to be considered in future surveillance and disease control activities.

**Funding source**

This work was supported by the Conselho Nacional de Desenvolvimento Científico e Tecnológico (CNPq) [grant number 306025/2019-1 to FCN]; and the São Paulo Research Foundation (FAPESP) [grant number 2017/10297-1 to CL].

**CRediT authorship contribution statement**

**Gleice Margarete de Souza Conceição:** Data curation, Formal analysis, Methodology, Software, Writing - original draft, Writing - review & editing. **Gerson Laurindo Barbosa:** Conceptualization, Investigation, Supervision, Writing - original draft, Writing - review & editing. **Camila Lorenz:** Investigation, Writing - original draft, Writing - review & editing. **Ana Carolina Dias Bocewicz:** Investigation, Writing - original draft, Writing - review & editing. **Lidia Maria Reis Santana:** Data curation, Investigation, Writing - original draft, Writing - review & editing. **Cristiano Corrêa de Azevedo Marques:** Conceptualization, Investigation, Supervision, Writing - original draft, Writing - review & editing. **Francisco Chiaravalloti-Neto:** Conceptualization, Investigation, Supervision, Writing - original draft, Writing - review & editing.

**References**

1. Bhatt S, Gething PW, Brady OJ, Messina JP, Farlow AW, Moyes CL & Myers MF. The global distribution and burden of dengue. Nature 2013; 496:504-507.

2. Schaber KL, Paz-Soldan VA, Morrison AC, Elson WHD, Rothman AL, Mores CN, Astete-Vega H, Scott TW, Waller LA, Kitron U, Elder JP, Barker CM, Perkins TA & Vazques-Prokopec GM. Dengue illness impacts daily human mobility patterns in Iquitos, Peru. PLoS Negl Trop Dis 2019; 13: e0007756. https://doi.org/10.1371/journal.pntd.0007756

3. Brady OJ, Hay SI. The global expansion of dengue: How *Aedes aegypti* mosquitoes enabled the first pandemic arbovirus. Annu Rev Entomol 2020; 65:91-208.

4. WHO. World Health Organization. 2020 https://www.who.int/health-topics/dengue-and-severe-dengue#tab=tab_1

5. Ministério da Saúde Brasil . Boletim Epidemiológico, vol. 47, n. 3. 2016. Available at: http://portalarquivos2.saude.gov.br/images/pdf/2016/janeiro/15/svs2016-540 be003-dengue-se52.pdf

6. Ministério da Saúde Brasil. Boletim Epidemiológico 13, vol. 50. 2019. Available at: http://portalarquivos2.saude.gov.br/images/pdf/2019/abril/30/2019-013-Monitoramento-dos-casos-de-arboviroses-urbanas-transmitidas-pelo-Aedes-publicacao.pdf

7. Achee NL, Gould F, Perkins TA, Reiner Jr RC, Morrison AC, Ritchie SA. A critical assessment of vector control for dengue prevention. PLoS Negl Trop Dis 2015; 9:e0003655.

8. Scott TW, Amerasinghe PH, Morrison AC, Lorenz LH, Clark GG, Strickman D, Kittayapong P & Edman JD. Longitudinal studies of *Aedes aegypti* (Diptera: Culicidae) in Thailand and Puerto Rico: blood feeding frequency. J Med Entomol 2000; 37:89-101.

9. Harrington LC, Scott TW, Lerdthusnee K, Coleman RC, Costero A, Clark GG, Jones JJ, Kitthawee S, Kittayapong P, Sithiprasasna R & Edman, J. D. Dispersal of the dengue vector *Aedes aegypti* within and between rural communities. Am J Trop Med Hyg 2005; 72: 209-220.

10. Chadee DD, Sutherland JM, Gilles JR. Diel sugar feeding and reproductive behaviours of *Aedes aegypti* mosquitoes in Trinidad: With implications for mass release of sterile mosquitoes. Acta Trop 2014; 132:86-S90.

11. Falcón-Lezama JA, Martínez-Vega RA, Kuri-Morales PA, Ramos-Castañeda J & Adams B. Day-to-day population movement and the management of dengue epidemics. B Math Biol 2016; 78: 2011-2033.

12. Vazquez-Prokopec GM, Montgomery BL, Horne P, Clennon JA & Ritchie SA. Combining contact tracing with targeted indoor residual spraying significantly reduces dengue transmission. Science Ad 2017; 3:e1602024.

13. Chiaravalloti-Neto F, Silva RA, Zini N, Silva GCD, Silva NS, Parra MCP, Dibo MR, Estofolete CF, Fávaro EA, Dutra KR, Mota MTO, Guimarães GF, Terzian ACB, Blangiardo M, Nogueira ML. Seroprevalence for dengue virus in a hyperendemic area and associated socioeconomic and demographic factors using a cross-sectional design and a geostatistical approach, state of São Paulo, Brazil. BMC Infect Dis 2019; 19:441.

14. Riley S. Large-scale spatial-transmission models of infectious disease. Science 2007; 316:1298-1301.

15. Stoddard ST, Morrison AC, Vazquez-Prokopec GM, Soldan VP, Kochel TJ, Kitron U, Elder JP & Scott TW. The role of human movement in the transmission of vector-borne pathogens. PLoS Negl Trop Dis 2009; 3:e481.

16. Kucharski AJ, Kwok KO, Wei VW, Cowling BJ, Read JM, Lessler J, Cummings DA & Riley S. The contribution of social behaviour to the transmission of influenza A in a human population. PLoS Pathog 2014; 10:e1004206.

17. Keeling MJ, Danon L, Vernon MC & House TA. Individual identity and movement networks for disease metapopulations. P Natl Acad Sci 2010; 107:8866-8870.

18. Arthur RF, Gurley ES, Salje H, Bloomfield LS & Jones JH. Contact structure, mobility, environmental impact and behaviour: the importance of social forces to infectious disease dynamics and disease ecology. Philos T R Soc B 2017; 372: 20160454.

19. Barbosa GL, Holcman MM, Pereira M, Gomes AHA & Wanderley DMV. Indicadores de infestação larvária e influência do porte populacional na transmissão de dengue no estado de São Paulo, Brasil: um estudo ecológico no período de 2007-2008. Epi Serv Sau 2012; 21:195-204.

20. Rodriguez-Morales AJ, Gallego V, Escalera-Antezana JP, Méndez CA, Zambrano LI, Franco-Paredes C & Risquez A. COVID-19 in Latin America: The implications of the first confirmed case in Brazil. Travel Med Infect Dis 2020;3:5-6

21. Lorenz C, Bocewicz ACD, de Azevedo Marques CC, Santana LMR, Chiaravalloti-Neto F, Gomes AHA & Barbosa GL. Have measures against COVID-19 helped to reduce dengue cases in Brazil? Travel Med Infect Dis 2020;4:3-4

22. SECRETARIA DE ESTADO DA SAÚDE. Centro de Vigilância Epidemiológica "Prof. Alexandre Vranjac". [Internet]. Available at: http://www.saude.sp.gov.br/cve-centro-de-vigilancia-epidemiologica-prof.-alexandre-vranjac/areas-de-vigilancia/doencas-de-transmissao-por-vetores-e-zoonoses/arboviroses-urbanas/dengue/dados-estatisticos. Accessed 16 Dec 2020.

23. FUNDAÇÃO SISTEMA ESTADUAL DE ANÁLISE DE DADOS – SEADE. Boletim Coronavírus Completo. Available at: https://www.seade.gov.br/coronavirus/. Accessed 16 Dec 2020.

24. Google LLC "Google COVID-19 Community Mobility Reports" Available at: https://www.google.com/covid19/mobility/ Acessed 15 Sep 2020.

25. Ramadona AL, Tozan Y, Lazuardi L, Rocklöv J. A combination of incidence data and mobility proxies from social media predicts the intra-urban spread of dengue in Yogyakarta, Indonesia. PLoS Negl Trop Dis 2019; 13:e0007298.

26. Stone CM, Schwab SR, Fonseca DM, Fefferman NH. Contrasting the value of targeted versus area-wide mosquito control scenarios to limit arbovirus transmission with human mobility patterns based on different tropical urban population centers. PLoS Negl Trop Dis 2019; 13: e0007479. https://doi.org/10.1371/journal.pntd.0007479

27. Zhang Y, Riera J, Ostrow K, Siddiqui S, de Silva H, Sarkar S, Fernando L & Gardner L. Modeling the relative role of human mobility, land-use and climate factors on dengue outbreak emergence in Sri Lanka. BMC Infect Dis 2020;20. https://doi.org/10.1186/s12879-020-05369-w

28. Soriano-Panos D, Arias-Castro A, Reyna-Lara HJ, Martinez S, Gomez-Gardenes J. "Vector-borne epidemics driven by human mobility" Phy Review Res 2020; 2:13312

29. Enduri MK, Jolad S. Dynamics of dengue disease with human and vector mobility . Spat Tem Epi 2018;25:57–66.

30. Gubler DJ. Dengue and dengue hemorrhagic fever. Clin Microbiol Rev 1998; 11: 480-496.

31. Chan M, Johansson MA. The incubation periods of dengue viruses. PloS One 2012; 7: e50972.

32. Yang HM, Macoris MLG, Galvani KC, Andrighetti MTM, Wanderley DMV. Assessing the effects of temperature on dengue transmission. Epi Infec 2009; 137: 1179-1187.

33. Rohani A, Wong YC, Zamre I, Lee HL, Zurainee MN. The effect of extrinsic incubation temperature on development of dengue serotype 2 and 4 viruses in *Aedes aegypti* (L.). Se Asian J Trop Med 2009; 40: 942-950.

34. Winokur OC, Main BJ, Nicholson J, Barker CM. Impact of temperature on the extrinsic incubation period of Zika virus in *Aedes aegypti*. PLoS Negl Trop Dis 2020; 14:e0008047.

35. Nacher M, Douine M, Gaillet M, Flamand C, Rousset D, Rousseau, Carles G. Simultaneous dengue and COVID-19 epidemics: Difficult days ahead?. PLoS Negl Trop Dis 2020; 14:e0008426.

36. Mascarenhas MDM, Batista FMDA, Rodrigues MTP, Barbosa ODAA, & Barros VC. Simultaneous occurrence of COVID-19 and dengue: what do the data show?. Cad Saude Publica 2020; 36: e00126520.

37. Kiang MV, Santillana M, Chen JT, Onnela JP, Krieger N, Monsen KE, Ekapirat N, Areechokchai D, Maude R, Buckee CO. Incorporating human mobility data improves forecasts of Dengue fever in Thailand 2020;5 doi: https://doi.org/10.1101/2020.07.22.20157966

38. Pessanha JEM, Caiaffa WT, César CC, & Proietti FA. Avaliação do Plano Nacional de Controle da Dengue. Cad Saude Publica 2009; 25:1637-1641. https://doi.org/10.1590/S0102-311X2009000700024

39. Teixeira MG, Costa MCN, Barreto F & Barreto ML. Dengue: twenty-five years since reemergence in Brazil. Cad Saude Publica 2009;25:S7-S18. https://doi.org/10.1590/S0102-311X2009001300002

40. Horstick O, Runge-Ranzinger S, Nathan MB, Kroeger A. Dengue vector-control services: how do they work? A systematic literature review and country case studies. Trans R Soc Trop Med Hyg 2010;104:379-86. doi:10.1016/j.trstmh.2009.07.027.

41. Figueiró AC, Sóter AP, Braga C, Hartz ZMA & Samico I. Análise da lógica de intervenção do Programa Nacional de Controle da Dengue. Rev Bras S Mat Inf 2010; s93-s106. https://doi.org/10.1590/S1519-38292010000500009

42. SECRETARIA DE ESTADO DA SAÚDE. Grupo Técnico Arboviroses. Diretrizes para a Prevenção e Controle das Arboviroses Urbanas no Estado de São Paulo. São Paulo, 2017. Available at: http://www.cvs.saude.sp.gov.br/up/Diretrizes%20controle%20arboviroses%20ESP%20-%202017.pdf. Accessed 23 Nov 2020.